\documentclass[aps,prb,a4paper,preprint,showpacs]{revtex4}
\usepackage{amsmath}
\usepackage{amsfonts}
\usepackage{amssymb}
\usepackage{slashbox}
\usepackage{graphicx}
\usepackage{amsbsy}

\begin{document}
\title{Quantum pumping with adiabatically modulated barriers in graphene }
\author{Rui Zhu\renewcommand{\thefootnote}{*} \footnote{Corresponding author. Electronic address:
rzhu@scut.edu.cn} and Huiming Chen  }
\address{Department of Physics, South China University of Technology,
Guangzhou 510641, People's Republic of China }

\begin{abstract}

We study the adiabatic quantum pumping characteristics in the
graphene modulated by two oscillating gate potentials out of phase.
The angular and energy dependence of the pumped current is
presented. The direction of the pumped current can be reversed when
a high barrier demonstrates stronger transparency than a low one,
which results from the Klein paradox. The underlying physics of the
pumping process is illuminated.

\end{abstract}


\maketitle

\narrowtext

Generally speaking, the transport of matter from low potential to
high potential excited by absorbing energy from the environment can
be described as a pump process. Its counterpart in quantum mechanics
involves coherent tunneling and quantum interference\cite{Ref1}.
Since the experimental realization of the quantum pump\cite{Ref1},
research on quantum charge and spin pumping has attracted increasing
interest\cite{Ref2, Ref3, Ref4, Ref5, Ref6, Ref7, Ref8, Ref9, Ref10,
Ref11, Ref12, Ref13, Ref14, Ref15, Ref16, Ref17, Ref18, Ref19}. The
current and noise properties in various quantum pump structures and
devices were investigated such as the magnetic-barrier-modulated two
dimensional electron gas\cite{Ref4}, mesoscopic one-dimensional
wire\cite{Ref6}, quantum-dot structures\cite{Ref5, Ref11, Ref12},
mesoscopic rings with Aharonov-Casher and Aharonov-Bohm
effect\cite{Ref7}, magnetic tunnel junctions\cite{Ref10}.
Correspondingly, theoretical techniques have been put forward for
the treatment of the quantum pumps [Refs.(2, 3, 18) and references
therein]. One of the most prominent is the scattering matrix
approach for ac transport, as detailed by Moskalets et
al.\cite{Ref3} who derived general expressions for the pump current,
heat flow, and the shot noise for an adiabatically driven quantum
pumps in the weak pumping limit. The pump current was found to vary
in a sinusoidal manner as a function of the phase difference between
the two oscillating potentials. It increases linearly with the
frequency in line with experimental finding. Recently, Park et
al.\cite{Ref5} obtained an expression for the admittance and the
current noise for a driven nanocapacitor in terms of the Floquet
scattering matrix and derived a nonequilibrium
fluctuation-dissipation relation. The effect of weak
electron-electron interaction on the noise was investigated by
Devillard et al.\cite{Ref6} using the scattering matrix renormalized
by interactions. Applying the Green¡¯s function approach, Wang et
al.\cite{Ref14, Ref15, Ref16} presented a nonperturbative theory for
the parametric quantum pump at arbitrary frequencies and pumping
strengths. Independently, Arrachea\cite{Ref17} presented a general
treatment based on nonequilibrium Green functions to study transport
phenomena in quantum pumps.

Work on graphene (discovered by Geim and his colleagues almost 5
years ago\cite{Ref21}) heated up quickly as researchers realized
that the material¡¯s two-dimensionality caused it to show unusual
quantum behaviors\cite{Ref22, Ref23, Ref24, Ref25, Ref26, Ref27,
Ref28, Ref29, Ref30, Ref31, Ref32, Ref33, Ref34, Ref35, Ref36}.
Graphene transistors\cite{Ref23, Ref24}, chemical
sensors\cite{Ref22}, electrodes\cite{Ref25}, scales\cite{Ref22} and
frequency generators\cite{Ref22} are some proposed potential
applications. Even though graphene is a low-energy system consisting
of a two dimensional honeycomb lattice of carbon atoms, its
quasiparticle excitations can be fully described by the
(2+1)-dimensional relativistic Dirac equation. The fundamental
property of the Dirac equation is often referred to as the
charge-conjugation symmetry. Klein paradox and chiral tunneling are
two major effects of it in graphene. A sufficiently strong
potential, being repulsive for electrons, is attractive for
positrons and results in positron states inside the barrier.
Matching between electron and positron wave functions across the
barrier leads to the high-probability tunneling described by the
Klein paradox. The chirality of quasiparticles requires conservation
of pseudospin (which is linked to different components of the same
spinor wave function and is parallel/antiparallel to the direction
of motion of electrons/holes) during tunneling and induces angular
anisotropy in transmission in single- and multi-barrier structures
in graphene. Due to its unusual structure, many extraordinary
behaviors of devices based on graphene have been observed, such as
the conductance minimum\cite{Ref21, Ref27}, resonant
tunneling\cite{Ref28}, the shot noise with the Fano factor close to
1/3\cite{Ref27, Ref29, Ref30, Ref31, Ref32}, the unconventional
Quantum Hall effect\cite{Ref33} and the edge-state-related quantum
spin Hall effect\cite{Ref34, Ref35}, and the electronic cooling
effect\cite{Ref36}. However, a graphene-based quantum pump has not
yet been considered in literature.

In this work, we focus on an adiabatic quantum pump device based on
a graphene monolayer modulated by two oscillating gate potentials.
The Klein paradox featured pump current is obtained and illuminated.

The crystal structure of undoped graphene layers is that of a
honeycomb lattice of covalent-bond carbon atoms. One valence
electron corresponds to one carbon atom and the structure is
composed of two sublattices, labeled by A and B. In the vicinity of
the ${\bf{K}}$ point and in the presence of a potential $U$, the
low-energy excitations of the gated graphene monolayer are described
by the two-dimensional (2D) Dirac equation
\begin{equation}
 v_F ({\bf{\sigma }} \cdot {\bf{\hat p}})   \Psi
= (E - U)\Psi,
\end{equation}
where the pseudospin matrix $\vec \sigma $ has components given by
Pauli's matrices and ${\bf{\hat p}} = (p_x ,p_y )$ is the momentum
operator. The ``speed of light" of the system is $v_{F}$, i.e., the
Fermi velocity ($v_F  \approx 10^6 $ m/s). The eigenstates of Eq.
(1) are two-component spinors $\Psi  = [\psi _A ,\psi _B ]^T $,
where $\psi _A $ and $\psi _B $ are the envelope functions
associated with the probability amplitudes at the respective
sublattice sites of the graphene sheet.

In the presence of a one-dimensional confining potential $U=U(x)$,
we attempt solutions of Eq. (1) in the form $\psi _A (x,y) = \phi _A
(x)e^{ik_y y} $ and $\psi _B (x,y) =i \phi _B (x)e^{ik_y y} $ due to
the translational invariance along the $y$ direction. The resulting
coupled, first-order differential equations read as
\begin{equation}
 d\phi _B /d\xi  + \beta \phi _B  = (\varepsilon  - u )\phi _A ,
\end{equation}
\begin{equation}
d\phi _A /d\xi  - \beta \phi _A  =  - (\varepsilon  - u )\phi _B .
\end{equation}
Here $\xi =x/L$, $\beta =k_y L$, $u = UL/\hbar v_F $, and
$\varepsilon =EL/\hbar v_F $ ($L$ is the width of the structure).
For a double-barrier structure with two square potentials of height
$U_1$ and $U_2$ respectively, Eqs. (2) and (3) admit solutions which
describe electron states confined across the well and propagating
along it. The transmission and reflection amplitude $t$ and $s$ is
determined by matching $\phi _{A}$ and $\phi _{B}$ at region
interfaces.

Following the standard scattering approach\cite{Ref2, Ref3, Ref37}
we introduce the fermionic creation and annihilation operators for
the carrier scattering states. The operator $ \hat a_{L }^\dag (E,
\theta ) $ or $ \hat a_{L} (E, \theta ) $ creates or annihilates
particles with total energy $E$ and incident angle $\theta $ in the
left lead, which are incident upon the sample. Analogously, we
define the creation $ \hat b_{L }^\dag (E, \theta ) $ and
annihilation $ \hat b_{L } (E, \theta ) $ operators for the outgoing
single-particle states. Considering a particular incident energy $E$
and incident angle $\theta$, the scattering matrix $s$ follows from
the relation
\begin{equation}
\left( {\begin{array}{*{20}c}
   {b_{L } }  \\
   {b_{R  } }  \\
\end{array}} \right) = \underbrace {\left( {\begin{array}{*{20}c}
   r & {t'}  \\
   t & {r'}  \\
\end{array}} \right)}_s\left( {\begin{array}{*{20}c}
   {a_{L  } }  \\
   {a_{R  } }  \\
\end{array}} \right),
\end{equation}
where, $t$ and $r$ are the scattering elements of incidence from the
left reservoir and $t'$ and $r'$ are those from the right reservoir.

The frequency of the potential modulation is small compared to the
characteristic times for traversal and reflection of electrons and
the pump is thus adiabatic. In this case one can employ an instant
scattering matrix approach, i.e. $s(t)$ depends only parametrically
on the time $t$. To realize a quantum pump one varies simultaneously
two system parameters, e.g. \cite{Ref2,Ref3}
\begin{equation}
\begin{array}{l}
 X_1 \left( t \right) = X_{10}  + X_{\omega ,1} e^{i\left( {\omega t - \varphi _1 } \right)}  + X_{\omega ,1} e^{ - i\left( {\omega t - \varphi _1 } \right)} , \\
 X_2 \left( t \right) = X_{20}  + X_{\omega ,2} e^{i\left( {\omega t - \varphi _2 } \right)}  + X_{\omega ,2} e^{ - i\left( {\omega t - \varphi _2 } \right)} . \\
 \end{array}
 \end{equation}
Here, $X_1$ and $X_2$ are measures for the two barrier heights $U_1$
and $U_2$, which can be modulated
 by applying two low-frequency ($\omega$)
alternating gate voltages.  $X_{\omega ,1} $ and $X_{\omega ,2} $
are the corresponding oscillating amplitudes with phases
$\varphi_{1/2}$;
  $X_{10}$ and $X_{20}$ are the static (equilibrium) components.

  As in the work of  Moskalets and B\"uttiker \cite{Ref3},
 in the weak pumping limit ($X_{\omega ,j}  \ll
X_{j0} $) and at zero temperature, the pump current could be
expressed in terms of the scattering matrix as follows.
\begin{equation} I_\alpha   =
\frac{{e\omega }}{{2\pi }}\sum\limits_{\beta j_1 j_2 } {X_{\omega
,j_1 } X_{\omega ,j_2 } \frac{{\partial s_{\alpha \beta }
}}{{\partial X_{j_1 } }}\frac{{\partial s_{\alpha \beta }^*
}}{{\partial X_{j_2 } }}2i\sin \left( {\varphi _{j_1 }  - \varphi
_{j_2 } } \right)}.
\end{equation}

The mechanisms of an adiabatic quantum pump can be demonstrated in a
mesoscopic system modulated by two oscillating barriers (see Fig.
1). We here consider a quantum pump without the effect of the Klein
paradox and look into the latter afterwards. To prominently picture
the charge flow driven process within a cyclic period, the two
potential barriers are modulated with a phase difference of $ \pi
/2$ in the manner of $U_1 =U_0 + U_{1 \omega } \sin t$ and $U_2 =U_0
+ U_{2 \omega } \sin (t+ \pi /2)$. Our discussion is within the
framework of the single electron approximation and coherent
tunneling. The Pauli principle is taken into account throughout the
pumping process. The Fermi energy of the two reservoirs and the
inner single-particle state energy are equalized to eliminate the
external bias and secure energy-conserved tunneling. As shown in
Fig. 1, the transmission strengths between one of the reservoirs and
the inner single-particle state are denoted by $t_1$-$t_4$. When $t
\in [0, \pi /2]$, $\sin t$ changes from 0 to 1 and $\sin (t+ \pi
/2)$ changes from 1 to 0. Considering the time-averaged effect, the
chance of $U_1 >U_2$ and $U_1 >U_2$ is equal. Therefore, the
probability of $t_1$ and $t_3$ balance out. The tunneling quantified
by $t_2$ and $t_4$ do not occur since the inner particle state is
not occupied. When $t \in [\pi /2,\pi ]$, $\sin t$ changes from 1 to
0 and $\sin (t+ \pi /2)$ changes from 0 to -1. $U_1
>U_2$ invariably holds in this time regime. The probability of $t_3$
prevails and a net particle flow is driven from the right reservoir
to the middle state. When $t \in [\pi , 3 \pi /2]$, $\sin t$ changes
from 0 to -1 and $\sin (t+ \pi /2)$ changes from -1 to 0. The
probability of $t_2$ and $t_4$ balance out and the tunneling
quantified by $t_1$ and $t_3$ are excluded from the Pauli principle.
No net time-averaged tunneling occur. When $t \in [3 \pi /2,2 \pi
]$, $\sin t$ changes from -1 to 0 and $\sin (t+ \pi /2)$ changes
from 0 to 1. $U_1$ maintains a lower height than $U_2$, which drives
the particle in the inner state to the left reservoir. Through one
whole pumping cycle, electrons (and positrons in graphene) are
pumped from the right reservoir to the left by absorbing energy from
the two oscillating sources. The tunneling is governed by quantum
coherence. In each period, the pumping process repeats and the
particles are driven continuously in the same direction as time
accumulates. Direction-reversed current can be obtained with
reversed phase difference of the two oscillating gates. The
direction of the pumped current is from the phase-leading gate to
the phase-lagged one without exception when we assume that higher
barriers admit smaller transmission probability. It can find
resemblance in its classical turnstile counterpart\cite{Ref38} with
the fore-opened gate admits transmission ahead of the later-opened
one driving currents in corresponding manner.

We now consider the pumped current in the graphene-based conductor.
In the numerical calculations, the parameters $U_{10}=U_{20}=100$
meV, $L=200$ nm, $U_{\omega , 1}=U_{\omega , 2}=0.1$ meV. The phase
difference of the two oscillating gate potentials $\Delta \phi =\phi
_1 -\phi _2$ is set at a constant value of $0.5$ (in radian). The
unimpeded penetration of quasiparticles (quasielectrons and
quasiholes) through high and wide potential barriers described by
Klein paradox is an exotic property of graphene resulting from its
particular double degenerate light-cone-like band structure.
Considering a graphene-based double-barrier structure, it is
possible to pump current from one reservoir to the other at zero
external bias by oscillating the potential barrier heights. To
demonstrate the Klein paradox induced pumping properties, we present
in Figs. 2 and 3 the angular and energy dependence of the pumped
current respectively with constant phase difference of the two gate
voltages. The positive pumped current is defined to be from the left
reservoir to the right. From Fig. 2, we observe zero net
electron/positron flow at normal incidence for all incident
energies. It is an effect of Klein tunneling. In terms of the
conservation of pseudospin, the barrier always remains perfectly
transparent for angles close to the normal incidence $\theta =0$.
Therefore, the adiabatic oscillation of the two gate potentials out
of phase would allow equal transmission rightward and leftward for
normal incidence throughout the period of their cyclic changes,
which generates no net current flow. Finite current flow is pumped
at the angles where pseudospin matching produces prominent
transmission. For $E=25$ meV, the pumped current is positive. For
$E=78$ meV, the pumped current is negative. For $E=92$ meV, the
pumped current curve flips back to the positive half plane of the
figure. For $E=98$ meV, the pumped current is positive for some
incident angles and negative for others. It is shown here the pumped
current can shift direction when the incident angle or energy change
for fixed $\Delta \phi$. The accumulated contribution of electrons
tunneling from different angles to the pumped current is shown in
Fig. 3. It can be seen that the flow direction of the angle-averaged
pumped current can change from rightward to leftward and reversely
as the energy changes. This is remarkable since in quantum pumps
based on usual mesoscopic nanoscale conductors considered in
literature the direction of the pumped current is determined by the
phase difference of the two oscillating parameters and remains
constant when the latter is fixed (see the preceding introduction).

The results can be interpreted by the mechanism of the pumping
process. Different from conventional tunnel barriers, the
transmission probability in graphene through a high barrier can
exceed that of a low barrier characterized by the Klein paradox. A
numerical comparison is given in Fig. 4. In a pumping cycle, the
phase-lagged gate can admit transmission in advance of the
phase-leading when a high barrier is more transparent than a low
one. Therefore the direction of the current flow can be reversed in
such conditions. Accordingly, the angle-averaged pumped current can
be in either direction even with the phase of the pump source fixed.

In summary, a quantum pump device involving the graphene-based
ballistic tunneling structure is investigated. For two independent
adiabatically modulated parameters of this device a finite net
charge current is transported. The physical mechanism of quantum
pumping is presented within the framework of single-particle
approximation and coherent tunneling. It is observed that the
direction of the pumped current can be reversed when a high barrier
demonstrates stronger transparency than a low one, which is a result
of the Klein paradox.

\clearpage

\begin{figure}[h]
 \caption{The tunneling
 scenario of an adiabatic quantum pump. The two shadowed blocks
 represent the left and right electron reservoirs respectively. The two barriers oscillate adiabatically in
 time. The middle bar indicates the
 single-particle state between the two barriers. The
 Fermi levels of the two reservoirs are the same and are leveled to the
 single-particle state within the conductor.
  $t_1$-$t_4$ indicate the transmission
 amplitudes between one of the two reservoirs and the middle
 single-particle state.}
\end{figure}

\begin{figure}[h]
 \caption{Angular dependence of the pumped current for different
 quasiparticle energy.
}
\end{figure}

\begin{figure}[h]
 \caption{Energy dependence of the angle-averaged pumped current.}
\end{figure}

\begin{figure}[h]
 \caption{Angular dependence of the static transmission probability
 of tunneling from the left reservoir to the right
 without gate potential oscillation
 for different heights of the right barrier. The height of the
 left
 barrier is fixed to be 100 meV. The Fermi energy of the reservoir is 72 meV.}
\end{figure}

\end{document}